\documentclass[aps,prl,reprint,superscriptaddress,nobibnotes,nofootinbib]{revtex4-2}
\usepackage[version=3]{mhchem}
\usepackage[T1]{fontenc}
\usepackage{graphicx}
\usepackage{epstopdf}
\usepackage{ifpdf}
\usepackage{dcolumn} 
\usepackage{bm}
\usepackage{amssymb}
\usepackage{amsmath}
\usepackage{epsfig}
\usepackage{upgreek}
\usepackage{textcomp}
\usepackage{gensymb}
\usepackage{natmove}
\usepackage{lmodern}
\usepackage{booktabs}
\usepackage{setspace}
\usepackage{adjustbox}
\usepackage{xcolor}
\usepackage{siunitx}
\usepackage{lipsum}
\usepackage{xcolor}
\usepackage{etoolbox}
\usepackage{multibib}
\makeatletter
\AtBeginDocument{%
  \@ifundefined{r@LastBibItem}{\global\@namedef{r@LastBibItem}{{0}{}}}{}%
}
\makeatother

\begin{document}
\title{Planar Ballistic Electron Emission Spectroscopy for Single-Shot Probing of Energy Barrier Inhomogeneity at Junction Interface}

\author{Jiwan Kim}  \affiliation{Department of Physics, Ulsan National Institute of Science and Technology (UNIST), Ulsan {\sl 44919}, Republic of Korea}

\author{Jaehyeong Jo}  \affiliation{Department of Physics, Ulsan National Institute of Science and Technology (UNIST), Ulsan {\sl 44919}, Republic of Korea}

\author{Jungjae Park} \affiliation{Department of Physics, Ulsan National Institute of Science and Technology (UNIST), Ulsan {\sl 44919}, Republic of Korea}

\author{Hyunjae Park}  \affiliation{Department of Physics, Ulsan National Institute of Science and Technology (UNIST), Ulsan {\sl 44919}, Republic of Korea}

\author{Eunseok Hyun}  \affiliation{Department of Physics, Ulsan National Institute of Science and Technology (UNIST), Ulsan {\sl 44919}, Republic of Korea}

\author{Jisang Lee}  \affiliation{Department of Physics, Ulsan National Institute of Science and Technology (UNIST), Ulsan {\sl 44919}, Republic of Korea}

\author{Sejin Oh}  \affiliation{Department of Physics, Ulsan National Institute of Science and Technology (UNIST), Ulsan {\sl 44919}, Republic of Korea}

\author{Kibog Park} \email{kibogpark@unist.ac.kr} \affiliation{Department of Physics, Ulsan National Institute of Science and Technology (UNIST), Ulsan {\sl 44919}, Republic of Korea} \affiliation{Department of Electrical Engineering, Ulsan National Institute of Science and Technology (UNIST), Ulsan {\sl 44919}, Republic of Korea}


\begin{abstract}
We propose an experimental methodology for probing the energy barrier inhomogeneity at the metal/semiconductor interface without the need for time-consuming microscopic survey. It is based on the known statistical nature of the interfacial energy barrier and the use of planar tunnel junction as an array of parallelly-connected ballistic electron emission microscopy (BEEM) tips. In order to analyze a lump of local BEEM signals, we incorporate the Tung model into the Bell-Kaiser theory. To validate our theoretical strategies, we investigate the interfacial energy barrier inhomogeneity of Pt/4H-SiC(0001) junction as a model system.
\end{abstract}


\maketitle
The energy barrier at the metal/semiconductor (MS) interface, well-known as the Schottky barrier (SB), is formed because the energy distribution of delocalized electronic states that contribute to the electrical conduction in bulk material is different on each side \cite{tung2014physics,sze}. This discontinuity is significantly affected by the interaction between orbitals of metal and semiconductor atoms at the interface, not only the difference in the work function of metal and the electron affinity of semiconductor \cite{tung2014physics,sze,tung2000chemical}. Thus, the SB is sensitive to the crystallography of the MS interface and can be different from region to region at the structually-inhomogeneous MS interface where the lattice translational symmetry in the transverse direction is broken \cite{tung1991transport,tung1992electron,tung1984nisi2}. Although we here focus mainly on the MS junction, it is apparent that the inhomogeneity of the interfacial energy barrier associated with crystallographic variations is common in any kind of heterojunction. \\
\indent Estimating inhomogeneity of the energy barrier is of great importance since it plays a crucial role in determining electrical properties of various technologically-relevant devices. Several such examples include the mobility degradation of the two-dimensional (2D) electron gas formed at the AlGaAs/GaAs interface in the low carrier density regime \cite{2deg}, the pinning in the type-I heterostructure-based 2D Wigner crystals \cite{wigner1,wigner2}, the robustness of interlayer charge transfer time regardless of the twisted angle of MoS\textsubscript{2}/WS\textsubscript{2} bilayer \cite{chargetransfer}, and the gradual turn-on of MS junction \cite{sze}. By virtue of the ballistic electron emission microscopy (BEEM) technique, which is a three-terminal variant of scanning tunneling microscopy, it has become possible to observe the local variation of interfacial energy barrier with high spatial resolution (on the order of 1-10 nm) \cite{bell1988observation}. Indeed, the previous BEEM measurements \cite{Im2001tung,talin1994} have revealed that the SB is spatially inhomogeneous and its barrier height distribution is Gaussian. Despite these astonishing achievements of BEEM technique, there are some drawbacks in using it to probe the barrier inhomogeneity: (1) The inevitable sudden jumps of BEEM current, related to the topography of somewhat rough metal surfaces, enforce repeating the same measurement to replace the jerky spectrum \cite{prietsch2001beem}. (2) Some supplemental measurement techniques, such as the scanning tunneling potentiometry \cite{pelz1990stm}, are needed to detect the tip-related artifacts. (3) The nanometer scale resolution ironically causes the difficulty of large area investigation. \\
\indent In this letter, we propose an experimental methodology, which we will call the planar ballistic electron emission spectroscopy (BEES), to probe the inhomogeneity of interfacial energy barrier at the MS junction using single-shot spectral measurements. The working principle of planar BEES is based on the known statistical nature of barrier height distribution and the metal-base vertical transistor structure \cite{Yi2010bees, transistorBEES} which functions equivalently to a dense group of BEEM tips connected in parallel. To quantitatively deconvolve planar BEES spectra, we developed a theoretical approach incorporating the Tung model \cite{tung1992electron} into the Bell-Kaiser theory \cite{bell1988observation}. With this, we can enable the rapid and quantitative probing of interface barrier inhomogeneity without tip-related issues and the size limitation of investigated area \cite{prietsch2001beem,pelz1990stm}. \\
\indent Figure 1(a) illustrates the band diagram and experimental setup of planar BEES measurements on the inhomogeneous MS interface, where the two, high and low, distinct SBs exist as shown in the inset of Fig. 1(a) and they do not affect each other \cite{tung1992electron}. Hot electrons are injected into the base layer through the tunnel barrier, and their energy $E$ distribution peaks around the Fermi-level $E_F$ of emitter (see the dashed box of Fig. 1(a)) due to the energy dependence of the tunneling probability $D(E)$ and the Fermi-Dirac distribution $f(E)$. This energy distribution peak is modulated by the base-emitter voltage $V_{BE}$, enabling the spectroscopic investigation. Some fraction of hot electrons injected into the base can travel ballistically through the thin metal base layer and can reach the MS interface. In the absence of thermal broadening of $f(E)$, the planar BEES signal, defined as the ratio of the collector current $I_C$ to the tunnel current $I_T$, is expected to increase abruptly at the $V_{BE}$ values corresponding to the low and high SBs, causing two peaks in the second derivative of planar BEES spectrum (see Fig. 1(b)) \cite{iets1,iets2,iets3,beemSD}. In a real system, the SB follows a Gaussian distribution, rather than a distribution composed of two discrete delta functions. Thus, a single broad second-derivative spectrum, made up of multiple second-derivative peaks, will be observed. The qualitative descriptions above allow us to draw two expectations: (1) The lump of second derivative peaks has a maximum at the $V_{BE}$ corresponding to the mean barrier height $\Phi_B^0$ since it is the most probable value in the Gaussian distribution. (2) There is a relationship between the standard deviation of SB distribution and the broadness of the lumped second derivative peak. \\
\begin{figure}[t] 
\begin{center}
\includegraphics[width=0.95\linewidth]{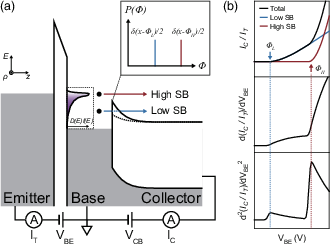}
\end{center}
\caption{(a) The band diagram of planar ballistic electron emission spectroscopy measurements for the inhomogeneous base/collector interface where the energy barrier consists of only two levels: low $\Phi_L$ and high $\Phi_H$ Schottky barrier. The corresponding probability density function $P(\Phi)$ is shown in the upper inset. The purple-colored distribution in the dashed box represents the energy distribution of hot electron flux. (b) The graphs on the right show the planar BEES spectrum along with its first and second derivative spectra.}
\end{figure}
\indent To demonstrate these two expectations quantitatively, we incorporate the Tung model \cite{tung1991transport} describing the potential landscape of inhomogeneous MS interface into the Bell-Kaiser theory \cite{bell1988observation} which is the standard theoretical tool for elucidating local BEES spectrum. For the typical inhomogeneous MS interface, the length scale of spatially varying SB is comparable to the depletion width $W$ of semiconductor \cite{tung1984nisi2,tung1992electron}. Thus, if there are low SB regions surrounded by the high SB, the potential can be \textit{pinched-off} \cite{tung1991transport}. It means that hot electrons must overcome the potential barrier $V_{th}$ within the depletion region, which is higher than the conduction band edge at the MS interface, to be collected as the substrate current. To include this potential \textit{pinch-off} in our theoretical approach, the spatial distribution of $V_{th}$ near the low SB "patch" is simplified using the Tung's point-dipole approximation \cite{tung1991transport}. By expanding up to the second-order at the saddle point of potential landscape near the low SB patch and truncating  unphysically high values \cite{Im2001tung}, $V_{th}$ around the patch can be approximated to be
\begin{align}
 \nonumber V_{th}(\gamma,\rho) \simeq \Phi _{B}^{0}-\bigg[&\gamma\left(\frac{V_{bb}}{\eta}\right )^{1/3}-\frac{9}{\gamma}\left(\frac{V_{bb}}{2W}\right)^{4/3}\rho^2\bigg] \\
&\times[1-\Theta(\rho-R_{patch}(\gamma))].
\end{align}
Here, $\gamma$ is the region parameter of low SB patch related to the strength of interfacial dipole moment \cite{tung1992electron} and $\rho$ is the distance from the center of low SB patch. $V_{bb}=\Phi_B^0+V_{CB}-(E_{C,bulk}-E_{F,s})$ is the band bending corresponding to the $\Phi_B^0$ where $V_{CB}$ is the collector-base voltage, $E_{C,bulk}$ is the conduction band minimum (CBM) of semiconductor in the neutral region, and $E_{F,s}$ is the Fermi-level of semiconductor. $R_{patch}(\gamma)=(\gamma/3)(16W^4/\eta V_{bb}^{3})^{1/6}$ is the radius of trucated low SB patch and $\eta=\varepsilon_s/qN_D^+$ where $q$ is the elementary charge and $N_D^+=N_D/[1+g_Dexp[(E_{F,s}-E_D)/k_BT]]$ is the ionized dopant concentration at temperature $T$ with the Boltzmann constant $k_B$. $\Theta$ is the Heaviside step function. With the assumption that the mutual interference among hot electron paths do not occur, the collector current through low SB patch can be expressed as 
\begin{equation}
I_{C,patch}(\gamma)= \int_{0}^{R_{patch}(\gamma)} J_{C,patch}(\gamma,\rho) \ 2\pi\rho \ d\rho
\end{equation}
where $J_{C, patch}(\gamma,\rho)$ is the collector current density contributed by the hot electrons going over the potential barrier located at distance $\rho$ from the center of the low SB patch characterized by $\gamma$. Although the transverse crystal momentum $k_\parallel$ is not a good quantum number for inhomogeneous MS interface due to the broken translational symmetry in the transverse direction, the Bell-Kaiser theory can be used here to describe $J_{C, patch}(\gamma,\rho)$ \cite{bell1988observation}. This is because the interface scattering that induces transitions between $k_\parallel$-conserved paths can only alter the relative magnitude of BEES current among several conduction band valleys located at different $k$-points in the Brillouin zone (BZ) or reduces the dependence of BEES spectra on the substrate surface orientation \cite{smith,schowalter1991elastic,LDbell1996,prietsch2}. Thus, we assume that the $E$ and $k_\parallel$ are conserved at the interface and the CBM of semiconductor is on-axis \cite{bell1988observation}. Then, $J_{C,patch}(\gamma,\rho)$ is given by
\begin{align}
\nonumber J_{C,patch}(\gamma,\rho)= RC &\int_{E_{min}(\gamma,\rho)}^{\infty} dE_\perp D(E_\perp) \\
&\times \int_{0}^{E_{max}(\gamma,\rho)} dE_\parallel f(E_\perp+E_\parallel)
\end{align}
\noindent where $R$ is a magnitude factor \cite{bell1988observation, Lee1991}, $E_\perp$ is the electron energy in the direction perpendicular to the interface and $E_\parallel$ is its parallel counterpart, $C=4\pi m_0 q/h^3$ with the Planck constant $h$ and the free electron mass $m_0$, $E_{min}(\gamma,\rho)=E_F-q(V_{BE}-V_{th}(\gamma,\rho))$, and $E_{max}(\gamma,\rho)=[m_t/(m_0-m_t)][E_\perp-E_F+q(V_{BE}-V_{th}(\gamma,\rho))]$ with the electron transverse effective mass of semiconductor $m_t$. \\
\indent Among a variety of MS interface morphology, we consider the case that the SB is $\Phi_B^0$ everywhere except in isolated patches with either lower or higher SBs \cite{tung1991transport,tung1992electron,Im2001tung}. By treating $\gamma$ as a Gaussian random variable \cite{tung1992electron}, we evaluate the planar BEES signal $I_C/I_T$ contributed by both the entire patches and the uniform background $\Phi_B^0$. Since the $I_C/I_T$ from high SB patches is turned-on at $V_{BE}$ values higher than $\Phi_B^0$, it can be neglected when $V_{BE}$ is not significantly higher than $\Phi_B^0$. Then, the $I_C/I_T$ becomes
\begin{widetext}
	\begin{align}
		\nonumber \frac{I_C}{I_T}=&R\ \frac{c_1\int_{0}^{\infty}d\gamma \frac{1}{\sigma \sqrt{2\pi }}exp\left(-\frac{\gamma^2}{2\sigma ^2}\right)\int_{0}^{R_{patch}(\gamma)}d\rho \ 2\pi\rho \int_{E_{min}(\gamma,\rho)}^{\infty}dE_\perp D(E_\perp)\int_{0}^{E_{max}(\gamma,\rho)}dE_\parallel f(E_\perp+E_\parallel)}{\int_{0}^{\infty}dE_\perp D(E_\perp)\int_{0}^{\infty}dE_\parallel[f(E_\perp+E_\parallel)-f(E_\perp+E_\parallel+qV_{BE})]}\\
&\hspace{11pc}+R(1-C_p)\frac{\int_{E_{min}(0,0)}^{\infty}dE_\perp D(E_\perp)\int_{0}^{E_{max}(0,0)}dE_\parallel f(E_\perp+E_\parallel)}{\int_{0}^{\infty}dE_\perp D(E_\perp)\int_{0}^{\infty}dE_\parallel[f(E_\perp+E_\parallel)-f(E_\perp+E_\parallel+qV_{BE})]}
	\end{align}
\end{widetext}
\noindent where $c_1$ is the number of low SB patches per unit area, $\sigma$ is the standard deviation of $\gamma$, and $C_p=\pi c_1\sigma^2(16W^4/\eta)^{1/3}/18V_{bb}$ is the areal coverage of low SB patches. The first term is the contribution from the low SB patches and the second term is the contribution from the uniform background $\Phi_B^0$. Equation (4) can be used to fit the second-derivative of experimentally-measured $I_C/I_T-V_{BE}$ curve by adjusting $\sigma$, $R$, and $c_1$ with the fixed $\Phi_B^0$ corresponding to the $V_{BE}$ value of maximum point in $d^2(I_C/I_T)/dV_{BE}^2-V_{BE}$ plot. Consequently, the barrier inhomogeneity is characterized by $\sigma$ and $c_1$. \\
\indent For deeper insights into the influence of barrier inhomogeneity on the planar BEES spectra, we calculated the planar BEES spectra and their second-derivative with different $c_1$ and $\sigma$ values using Eq. (4), as shown in Figs. 2(a)-(d). We normalized each spectrum by its maximum value for comparison. The normalized $I_C/I_T$ in Fig. 2(a) appears to be turned on at lower $V_{BE}$ as $\sigma$ increases. It makes sense because the likelihood of including even lower SBs in the MS interface increases as $\sigma$ increases. In contrast, the normalized $I_C/I_T$ in Fig. 2(b) is observed to be turned on at similar $V_{BE}$ regardless of $C_p$ but exhibits more gradual turn-on characteristics as $C_p$ increases. It is because $C_p$ alters only the ratio of $I_C/I_T$ contributed by low SB patches to that contributed by uniform background $\Phi_B^0$. The changes driven by $\sigma$ and $C_p$ are more clearly reflected in the normalized $d^2(I_C/I_T)/dV_{BE}^2-V_{BE}$ plot. The second-derivative spectrum in Fig. 2(c) becomes broadened for $V_{BE}$ below $\Phi_B^0$ as $\sigma$ increases, which is attributed to the overlap of multiple second-derivative peaks associated with low SBs. In the case of $C_p$, the normalized $d^2(I_C/I_T)/dV_{BE}^2$ at the sharp bend ($V_{BE}\sim$1.6 V) in Fig. 2(d) increases progressively with $C_p$ being increased since the $I_C/I_T$ contributed by the uniform background $\Phi_B^0$ is turned on at that point. These apparently different roles of $\sigma$ and $C_p$ in the BEES spectra can serve to mitigate the risk of overfitting. As predicted previously, the peak position in Fig. 2(d) is almost invariant even if the MS interface is perfectly covered by low SB patches. Also, this peak position can be slightly higher than the true $\Phi_B^0$ due to the thermal broadening of $f(E)$, which is negligible at low temperatures. \\
\begin{figure}[b] 
\begin{center}
\includegraphics[width=0.95\linewidth]{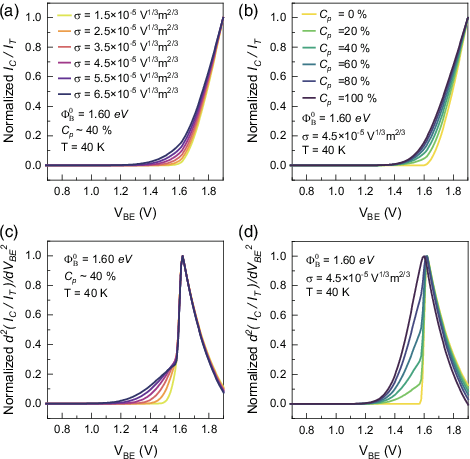}
\end{center}
\caption{The planar BEES spectra are calculated by Eq. (4) with different (a) standard deviation of region parameter $\sigma$ and (b) areal coverage of low SB patches $C_p$. (c) The second derivative of (a) and (d) that of (b) are also given. All spectra are normalized by their maximum value for comparison.}
\end{figure}
\begin{figure*}[t] 
\begin{center}
\includegraphics[width=1.0\linewidth]{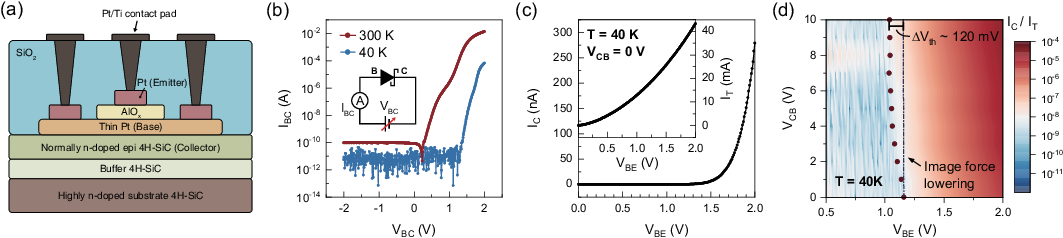}
\end{center}
\caption{(a) The schematics of the planar BEES architecture in the form of hot electron transistor used in our experiments. (b) The current-voltage $I-V$ characteristics of Pt/4H-SiC Schottky junction. (c) The collector current $I_C$ of Pt/AlO\textsubscript{x}/Pt/4H-SiC hot electron transistor measured at 40 K. The inset data represents the tunneling current $I_T$ of the Pt/AlO\textsubscript{x}/Pt tunnel junction. (d) The $V_{BE}$ and $V_{CB}$ scan of planar BEES signal $I_C/I_T$ measured at 40 K. The dark red dots represent the threshold voltage extracted from the linear extrapolation in $(I_C/I_T)^{1/2}-V_{BE}$ plot.}
\end{figure*}
\indent As a model system of planar BEES, we fabricated a Pt (50 nm)/AlO\textsubscript{x} ($\sim$1-2 nm)/Pt (4 nm)/4H-SiC hot electron transistor and mounted it on a cryogenic system (see Supplemental Material for details). Figure. 3(a) shows a schematic cross-sectional view of the fabricated device. The 4H-SiC is composed of a 4$\degree$ miscut Si-face substrate heavily n-doped with a nitrogen density of $\sim$10\textsuperscript{19} cm\textsuperscript{-3}, a buffer layer with nitrogen doping concentration $N_D$ of $\sim$10\textsuperscript{18} cm\textsuperscript{-3}, and a 15 $\mu$m epitaxial layer grown on the buffer layer with $N_D\sim$ 2$\times$10\textsuperscript{15} cm\textsuperscript{-3}. The 4H-SiC was prepared by going through ultrasonication-assisted solvent cleaning followed by HF dipping. The Pt base layer with an area of 600 × 1800 $\mu$m\textsuperscript{2} was immediately deposited on the 4H-SiC collector by using e-beam evaporation. The AlO\textsubscript{x} tunnel barrier was formed by the \textit{in-situ} oxidation in pure O\textsubscript{2} ($\sim99.999 \ \%$) after a thin Al film being thermally evaporated. The Pt emitter layer was then deposited onto the AlO\textsubscript{x} barrier using e-beam evaporation through a shadow mask with a 50 $\mu$m-diameter circular opening. The Pt film was also deposited simultaneously onto the thin Pt base to protect it during the VIA formation. The standard device fabrication processes were used for the remaining structures \cite{supplementary}. \\
\indent In order to investigate the Pt/4H-SiC interface characteristics and verify the functionality of hot electron transistor, we measured the current-voltage (I-V) curve of Pt/4H-SiC Schottky junction (Fig. 3(b)) and the collector current $I_C$ of hot electron transistor (Fig. 3(c)). The hump observed in the I-V curve of Pt/4H-SiC junction at 300 K implies the presence of crystal defects on the surface of 4H-SiC wafer, such as stacking faults, polytype inclusions, and epilayer pits \cite{tung1992electron,Im2001tung, Skromme2000ebic}. Even with those discrete crystal defects, it is very likely that the planar BEES spectra remain unaffected by these \textit{extrinsic} defects due to their low surface coverage \cite{Im2001tung}. Thus, we consider only the \textit{intrinsic} barrier inhomogeneity due to the polycrystalline metal film \cite{tung1991transport, tung1984nisi2, Im2001tung}. The $I_C$ of hot electron transistor measured at 40 K in Fig. 3(c), obtained by averaging 100 curves to improve the signal-to-noise ratio, shows the typical $V_{BE}$ dependence with the turn-on behavior \cite{transistorBEES}. It is sufficiently higher than the reverse current of Pt/4H-SiC junction at 40 K, indicating that the tunnel oxide is not shorted and hot electrons are injected successfully into the Pt base through the AlO\textsubscript{x} tunnel barrier. \\
\indent The color map of $I_C/I_T$ at 40 K was taken by scanning $V_{BE}$ and $V_{CB}$, as shown in Fig. 3(d). Each $I_C/I_T-V_{BE}$ curve in the color map are averaged over 100 curves. The effective threshold $V_{th}^{eff}$ was determined by the linear extrapolation in the $(I_C/I_T)^{1/2}-V_{BE}$ plot, which we call the Bell-Kiaser plot \cite{supplementary, bell1988observation}. The $V_{th}^{eff}$ depicted as the dark red dot in the color map is found to decrease as the $V_{CB}$ increases. This $V_{th}^{eff}$ decrease can be contributed from both image force lowering and barrier inhomogeneity. The image force lowering originates from the attractive charges induced on the metal surface which lower the energy barrier seen by the electrons approaching to the interface \cite{sze}. In the case of barrier inhomogeneity, the \textit{pinch-off} phenomenon of energy band profile in low-barrier areas is responsible for the $V_{th}^{eff}$ dependence on $V_{CB}$ \cite{tung1992electron}. In order to check which one is a dominant factor, we conducted the numerical calculations \cite{supplementary} to quantify the image force lowering $\Delta\Phi_{IFL}=[qN_D^+V_{bb}/8\pi^2\varepsilon_s^3]^{1/4}$. The charge neutrality condition $p-n+N_D^+=0$ with $n$ ($p$) being the electron (hole) density and the temperature dependence of $N_D^+$ were utilized to calculate the $E_{C,bulk}-E_{F,s}$ iteratively \cite{sze, FDintegral}. The calculated $\Delta\Phi_{IFL}$ is represented as the dashed line in Fig. 3(d) which hardly exhibits the $V_{CB}$ dependence due to the carrier freeze-out. Hence, we can conclude that the large shift of $V_{th}^{eff}$ ($\sim$120 mV) at 40 K manifests the dominancy of barrier inhomogeneity.
\\   
\begin{figure}[t] 
\begin{center}
\includegraphics[width=0.7\linewidth]{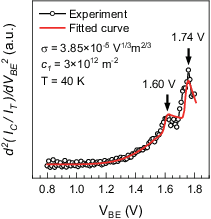}
\end{center}
\caption{The $d^2(I_C/I_T)/dV_{BE}^2-V_{BE}$ data with the fitted curve to Eq.(4) assuming two thresholds.}
\end{figure}
\indent The experimental $d^2(I_C/I_T)/dV_{BE}^2-V_{BE}$ curve for Pt/AlO\textsubscript{x}/Pt/4H-SiC hot electron transistor is shown in Fig. 4. Instead of using the second-harmonic detection method \cite{iets1,iets2,iets3}, it was obtained by smoothing with the 9 points cubic convolution and the differentiation of spectrum \cite{SG}. Two peaks are observed at 1.60 V and 1.74 V in the second-derivative spectrum, corresponding to the two lowest CBM of 4H-SiC located at the $M$ point in BZ, as reported in \cite{kaczer}. Using these two values, we fitted the experimental $d^2(I_C/I_T)/dV_{BE}^2$–$V_{BE}$ data up to $\sim$0.25 V above the lowest CBM of 4H-SiC to Eq. (4) with four parameters: two attenuation coefficients, $\sigma$, and $c_1$ \cite{kaczer, smith}. The $m_t$ value for each CBM used in \cite{kaczer} were adopted in this task. The tunnel barrier height of $\phi=3.7$ eV and thickness of $d_{t}=0.81$ nm were used to reproduce a tunnel current in the inset of Fig. 3(c) \cite{tunnel}. As shown in Fig. 4, our theory shows the satisfactory agreement with the experiment, and predicts the standard deviation of SB $\sigma_\Phi=\sqrt{C_p}(V_{bb}/\eta)^{1/3}\sigma=156.7$ mV. The expected sharp bending in the second derivative spectrum was not observed. This is likely due to the fact that the background $\Phi_B^0$ of a real MS interface is indeed not perfectly uniform, as shown in the previous BEEM study \cite{Im2001tung}. \\
\indent In summary, we suggest that the planar BEES can be used to determine the barrier inhomogeneity with the Bell-Kaiser theory incorporating the Tung model. Two crucial conclusions can be drawn: (1) The peak point in the second derivative of planar BEES spectra corresponds to $\Phi_B^0$ and remains nearly unaffected by barrier inhomogeneity. (2) The second derivative of planar BEES spectra broadens as $\sigma$ increases. With these discriminative aspects of planar BEES, the barrier inhomogeneity can be probed in convenient and fast manners without performing microscopic investigation such as BEEM. Our approach can extend its usage also to probe the barrier inhomogeneities of other heterojunction interfaces if the inhomogeneous interface possesses the Gaussian nature. \\
\indent This work is supported by the National Research Foundation of Korea (NRF) and Institute for Information and Communication Technology Planning and Evaluation(IITP) funded by Ministry of Science and ICT (NRF-2023R1A2C1006519, RS-2023-00227854). This work has also benefited from the use of the facilities at UNIST Central Research Facilities.

\clearpage         
\onecolumngrid

\setcounter{secnumdepth}{3}
\pagenumbering{arabic}   
\setcounter{page}{1}
\renewcommand{\thesection}{S\arabic{section}}
\setcounter{equation}{0}
\renewcommand{\theequation}{S\arabic{equation}}
\setcounter{figure}{0}
\renewcommand{\thefigure}{S\arabic{figure}}
\renewcommand{\thetable}{S\arabic{table}}

\begin{center}
  \textbf{\Large Supplemental Material: \\ 
  Planar Ballistic Electron Emission Spectroscopy for Single-Shot Probing of Energy Barrier Inhomogeneity at Junction Interface}\\[12pt]
  \normalsize
  Jiwan Kim,$^{1}$ Jaehyeong Jo,$^{1}$ Jungjae Park,$^{1}$ Hyunjae Park,$^{1}$\\
  Eunseok Hyun,$^{1}$ Jisang Lee,$^{1}$ Sejin Oh,$^{1}$ and Kibog Park$^{1,2,*}$\\[3pt]
  \small
  \textit{$^{1}$Department of Physics, Ulsan National Institute of Science and Technology (UNIST), Ulsan 44919, Republic of Korea}\\
  \textit{$^{2}$Department of Electrical Engineering, Ulsan National Institute of\\ Science and Technology (UNIST), Ulsan 44919, Republic of Korea}
\end{center}
\begingroup
  \renewcommand\thefootnote{\fnsymbol{footnote}}
  \footnotetext[1]{kibogpark@unist.ac.kr}
\endgroup
\vspace{3em}

\makeatletter
\patchcmd{\section}{\centering}{\raggedright}{}{}
\makeatother

\section{Device fabrication processes and experimental setup}
\begin{figure}[b] 
\begin{center}
\includegraphics[width=0.8\linewidth]{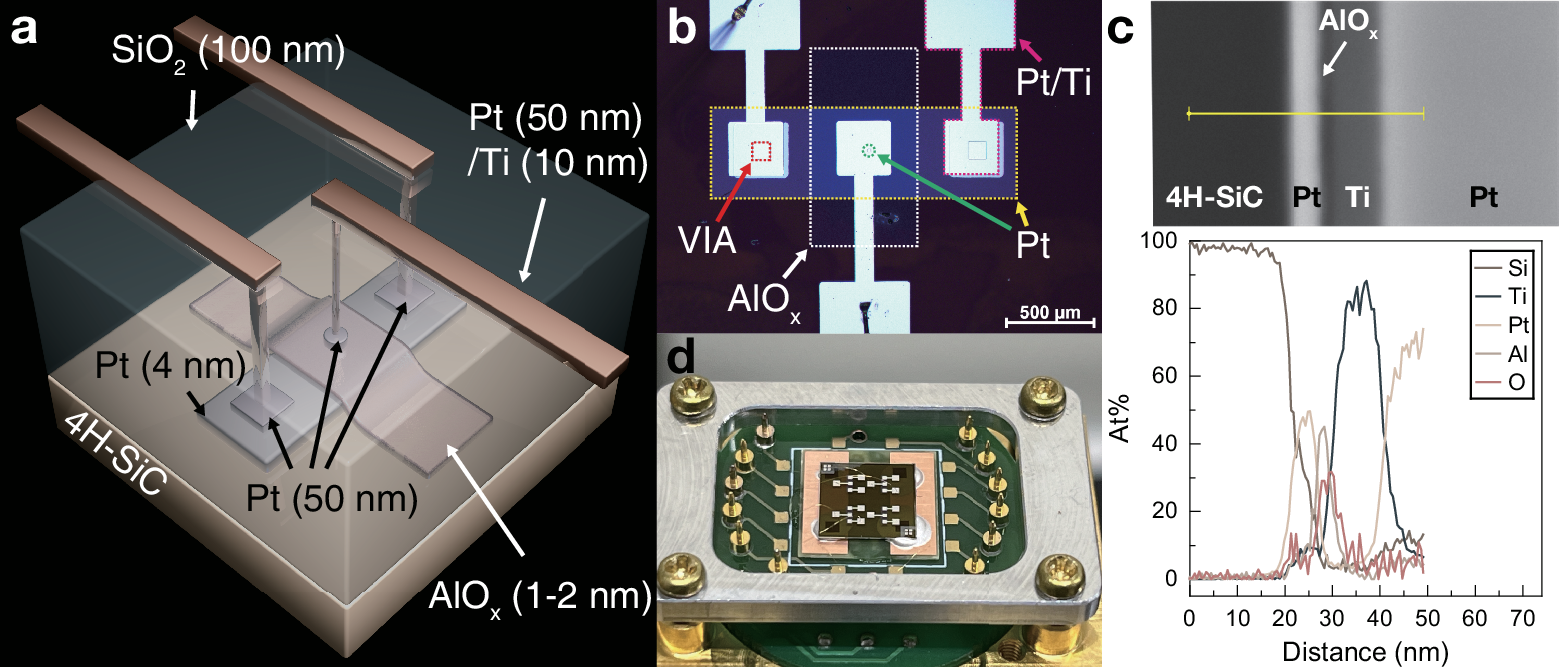}
\end{center}
\caption{(a) The schematic image and (b) the optical microscope image of the Pt (50 nm)/AlO\textsubscript{x} ($\sim$1-2 nm)/Pt (4 nm)/4H-SiC hot electron transistor. (c) The TEM image of the Pt/Ti/AlO\textsubscript{x}/Pt tunnel junction on the 4H-SiC with its EDS spctra. (d) The camera image of fabricated device with the PCB chip mounted on the cryogenic system.}
\end{figure}
We fabricated a Pt (50 nm)/AlO\textsubscript{x} ($\sim$1-2 nm)/Pt (4 nm)/4H-SiC hot electron transistor with a VIA structure (see Figs. S1(a) and (b)). The 4H-SiC sample, with a 100 nm protective SiO\textsubscript{2} layer deposited on the epilayer by a plasma enhanced chemical vapor deposition, was cleaved to 7 mm $\times$ 7 mm size. The sample was cleaned by the ultrasonic treatment with the acetone and methanol sequentially, followed by a 5-second dip in a 100:1 dilute hydrogen fluoride (dHF) solution to remove the native oxide from the back side of the substrate. With minimal air exposure, a 100 nm nickel layer was deposited on the back side of the substrate through a shadow mask using the electron beam (e-beam) evaporation, and the nickel silicide for ohmic contact was formed through the rapid thermal annealing at 1000 $^{\circ}C$ for 90 s. A 100 nm protective SiO\textsubscript{2} layer was removed by immersing our sample in 49$\%$ hydrogen fluoride for one minute. Immediately afterward, a 4 nm base layer of Pt, with an area of 600$\times$1800 $\mu$m\textsuperscript{2}, was deposited using the e-beam evaporation through a shadow mask. A $\sim$1-2 nm tunnel barrier oxide AlO\textsubscript{x}, with an area of 1100$\times$600 $\mu$m\textsuperscript{2}, was formed by depositing an Al thin film through a shadow mask, followed by oxidation in 140 torr of a pure O\textsubscript{2} ($\sim$99.999 $\%$) for 24 hours. The 50 sccm of pure O\textsubscript{2} flowed into the chamber to reach 140 torr from the base pressure of 5$\times$10\textsuperscript{-7} torr. To verify the thickness and quality of AlO\textsubscript{x} film, the test sample Pt/Ti/AlO\textsubscript{x}/Pt on the 4H-SiC was prepared by the same fabrication conditions. As shown in Fig. S1(c), the transmission electron microscopy (TEM) image and the energy-dispersive X-ray spectroscopy (EDS) of the test sample indicate that the AlO\textsubscript{x} film has a thickness of $\sim$1-2 nm and was sufficiently oxidized. A 50 nm thick Pt layer was deposited as the emitter of a 50 $\mu$m diameter dot and a base contact layer of 300$\times$300 $\mu$m\textsuperscript{2}, using the e-beam evaporation through a shadow mask. The VIA structure was formed by the radio frequency (RF) sputtering of a 10-inch SiO\textsubscript{2} target (Ar gas flow of 150 sccm, 15 mtorr, 800 W, 100 s devided into the 4 steps), the standard photolithography, and dipping in the 100:1 dHF for 80 s. In order to remove the photoresist, the ultrasonic solvent cleaning and the plasma treatment (O2 gas flow of 300 sccm, 50 Pa, 200 W, 15 min) was conducted. The contact metal Pt (50 nm)/Ti (10 nm) was deposited by the e-beam evaporation with a shadow mask. Our sample was attached to 200 $\mu$m thick NP-140TL (10$\times$10 mm\textsuperscript{2}) with the two stripe-shaped copper electrodes by using the silver paste, and Apiezon N grease (see Fig. S1(d)). Lastly, this sample was affixed to the PCB chip using GE varnish and Apiezon N grease. Each electrode was wire-bonded to the contact pads on the PCB chip. \\
\indent The fabricated devices were mounted on our cryogenic measurement system. After checking the electrical connections between our devices and BNC cables, the sample holder was enclosed in the chamber. The rotary pump was turned on for an hour. The system was cooled by the cryogenic helium compressor (CNA-11). Before the temperature went below 210 K, the vacuum valve of the chamber was closed to prevent the backflow. After reaching low temperatures of around 2.67$\pm$0.05 K, the temperature was controlled and heated with the proportional–integral–derivative system in the temperature controller (LakeShore 335). After stabilizing the temperature at 40.025$\pm$0.005 K, electrical measurements were performed with the source measure units (Keithely 2636B).
\begin{figure}[b] 
\begin{center}
\includegraphics[width=1.0\linewidth]{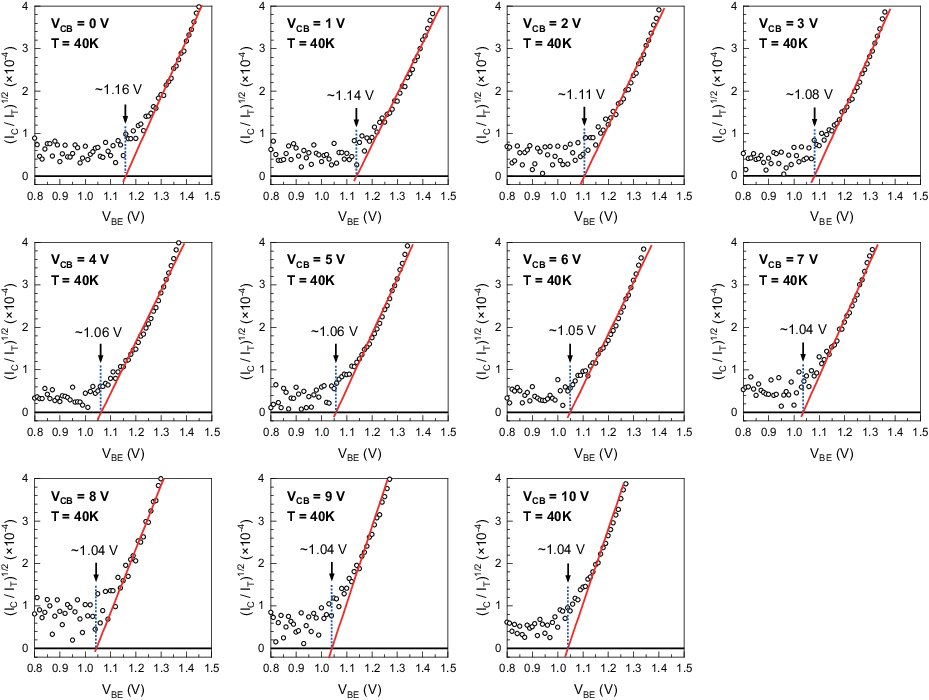}
\end{center}
\caption{The illustration of the threshold voltage $V_{th}^{eff}$ extraction method for the Pt/AlO\textsubscript{x}/Pt/4H-SiC hot electron transistor. The circle data points represent the planar BEES spectra in the Bell-Kiaser plot for different collector-base voltages $V_{CB}$. The red lines are the fitted line, where each planar BEES spectrum is fitted near the threshold.}
\end{figure}
\section{Extraction of $V_{th}^{eff}$ from Bell-Kaiser plot}
In accordance with the Bell-Kaiser theory \cite{bell1988observation2}, the planar BEES spectrum of the homogeneous metal/semiconductor (MS) interface follows the square-law near the threshold. Therefore, the effective threshold voltage $V_{th}^{eff}$ of the planar BEES spectrum was extracted by determining the x-intercept from the linear extrapolation of the Bell-Kiaser plot near the threshold, as shown in Fig. S2. With the assumption that the barrier inhomogeneity cannot affect the planar BEES spectrum, the collector-base voltage $V_{CB}$ dependence of the $V_{th}^{eff}$ must be explained by the image force lowering. By inductive reasoning, if image force lowering fails to describe the $V_{CB}$ dependence of the $V_{th}^{eff}$, it suggests that the planar BEES spectrum is influenced by barrier inhomogeneity. \\
\section{Numerical calculations of image force lowering}
We quantify the image force lowering $\Delta\Phi_{IFL}=[qN_D^+V_{bb}/8\pi^2\varepsilon_s^3]^{1/4}$ where $q$ is th elementary charge, $N_D^+$ is the ionized dopant concentration, $\Phi_B^0$ is the mean Schottky barrier (SB), $V_{CB}$ is the collector-base voltage, $E_{C,bulk}$ is the conduction band minima (CBM) of semiconductor in the neutral region, $E_{F,s}$ is the Fermi-level of semiconductor, $V_{bb}=\Phi_B^0+V_{CB}-(E_{C,bulk}-E_{F,s})$ is the band bending, and $\varepsilon_s$ is the permittivity of semiconductor. Since the Fermi-level position in the neutral region of the semiconductor is the only unknown variable, we solve a system of equations, including (\textit{i}) the charge neutrality condition and (\textit{ii}) the temperature dependence of the ionized dopant concentration \cite{sze2}, by using a \textit{vpasolve}() function in the MATLAB:

\begin{equation}
-n+p+N_D^+=0
\end{equation}
\\
\noindent where n (p) is the electron (hole) carrier density.

\begin{equation}
N_D^+=\frac{N_D}{1+g_Dexp[(E_D-E_{F,s})/k_BT]}
\end{equation}
\\[12pt]
\noindent where $N_D$ is the dopant concentration, $g_D$ is the ground-state degeneracy of the donor level, $E_D$ is the donor level, $k_B$ is the Boltzmann constant, and $T$ is the temperature. From Eqs. (S1) and (S2), the equation for $V_n=E_{C,bulk}-E_{F,s}$ can be derived \cite{FDintegral2}.

\begin{equation}
\frac{2}{\sqrt{\pi}}\left[N_CF_{1/2}(-V_n/k_BT)+N_VF_{1/2}((V_n-E_g)/k_BT)\right]=\frac{N_D}{1+g_Dexp[-(E_{C,bulk}-E_D)/k_BT]exp[V_n/k_BT]}
\end{equation}
\\[12pt]
where $F_{1/2}(x)$ is the Fermi-Dirac integral, $N_C$ ($N_V$) is the effective density of states in the conduction (valence) band, and $E_g$ is the band gap of semiconductor. We used the material parameters for the 4H-SiC in Table S1. From Eq. (S3), we can find the $V_n$, and calculate the $V_{bb}$ and the $N_D^+$ to quantify the $\Delta\Phi_{IFL}$.

\begin{table}[htbp]
  \centering
  \caption{Material parameters for the 4H-SiC}
    \begin{tabular}{lccccc}
        \toprule
          Material Parameters                & Values             \\
        \midrule
          $N_C$ (cm\textsuperscript{-3})                     & $(1.68\times 10^{19})(T/300)^{1.5}$                  \\
          $N_V$ (cm\textsuperscript{-3})               & $(2.49\times 10^{19})(T/300)^{1.5}$           \\
          $E_g$ (eV)      &$3.23-(6.5\times 10^{-4})T^2/[T+1300]$             \\
          $N_D$ (cm\textsuperscript{-3})      & 2$\times$10\textsuperscript{15}                      \\
          $E_{C,bulk}-E_D$ (eV)   & 0.06                   \\
          $\varepsilon_s$ (F/m)                  &9.66$\varepsilon_0$  \\
        \bottomrule
    \end{tabular}
\end{table}

\section{Strong $V_{CB}$ dependence of $V_{th}^{eff}$ in planar ballistic electron emission spectroscopy (BEES) spectrum}
With the combination of the Bell-Kaiser theory \cite{bell1988observation2} and the R.Tung's theory \cite{tung1992electron2,Im2001tung2}, the ballistic-electron-emission-spectroscopy (BEES) current \textit{I\textsubscript{C} / I\textsubscript{T}} can be represented as the following:

\begin{align}
	\nonumber  \frac{I_C}{I_T}=&R\ \frac{\int_{0}^{\infty}dx \frac{c_1}{\sigma \sqrt{2\pi }}exp\left(-\frac{x^2}{2\sigma ^2}\right)\int_{0}^{R(x)}d\rho \ 2\pi\rho \int_{E_{min}(x,\rho)}^{\infty}dE_\perp D(E_\perp)\int_{0}^{E_{max}(x,\rho)}dE_\parallel f(E_\perp+E_\parallel)}{\int_{0}^{\infty}dE_\perp D(E_\perp)\int_{0}^{\infty}dE_\parallel[f(E_\perp+E_\parallel)-f(E_\perp+E_\parallel+qV_{BE})]}&\\[11pt]
&\hspace{10pc}+R\ \frac{(1-C_p)\int_{E_{min}(0,0)}^{\infty}dE_\perp D(E_\perp)\int_{0}^{E_{max}(0,0)}dE_\parallel f(E_\perp+E_\parallel)}{\int_{0}^{\infty}dE_\perp D(E_\perp)\int_{0}^{\infty}dE_\parallel[f(E_\perp+E_\parallel)-f(E_\perp+E_\parallel+qV_{BE})]}&
\end{align}
\\[12pt]
\noindent where $R$ is a measure of attenuation due to inelastic scattering in metal base layer including the quantum mechanical reflection and the phonon scatterng at the MS interface \cite{Lee19912}, $\sigma$ is the standard deviation of region parameter, $c_1$ is the number of circular patches per unit area, $W$ is the depletion width corresponding to uniform SB of $\Phi_{B}^{0}$, $\eta=\frac{\varepsilon_r\varepsilon_0}{qN_D^+}$, $R(x)=\frac{x}{3}\left(\frac{16W^4}{\eta V_{bb}^3}\right)^{1/6}$ is the trucated radius of circular patch \cite{Im2001tung}, $V_{th}(x,\rho)=\Phi _{B}^{0}-\left(\frac{V_{bb}}{\eta}\right )^{1/3}x+\frac{9}{x}\left(\frac{V_{bb}}{2W}\right)^{4/3}\rho^2$ corresponds to the threshold voltage of low SB patches \cite{tung1992electron2,Im2001tung2}, $V_{BE}$ is the base-emitter voltage, $E_F$ is the Fermi-level of the emitter, $f(E)=1/\left[1+exp((E-E_F)/k_BT)\right]$ is the Fermi-Dirac distribution, $D(E)$ is the tunneling probability across the tunnel oxide barrier \cite{simmons1963generalized}, $E_{min}(x,\rho)=E_{F}-q[V_{BE}-V_{th}(x,\rho)]$ \cite{bell1988observation2}, $m_0$ is the free electrom mass, $m_t$ is the transverse effective electron mass in the semiconductor, $E_{max}(x,\rho)=[E_\perp-E_F+q(V_{BE}-V_{th}(x,\rho))]m_t/(m_0-m_t)$ \cite{bell1988observation2}, and $C_p=\frac{\pi c_1\sigma^2}{18V_{bb}}\left(\frac{16W^4}{\eta}\right)^{1/3}$ is the areal coverage of low SB patches \cite{tung1992electron2,Im2001tung2}. The first (second) term in Eq. (S4) denotes the contributions of electrical current from the low SB patches (uniform background except the region of all circular patches). Because the electrical current across the high SB patches cannot significantly affect to the BEES turn-on characteristics at the near-threshold voltage region, we neglect the contributions of electrical current from the high SB patches. $R$ is taken to be same for low SB patches and uniform background, since the hot-electron attenuation lengths in metals are almost energy-independent for $E-E_F<2 eV$ \cite{bell1988observation2}. With the assumption that the Schottky diode consists of many circular patches, we will also ignore the second term in Eq. (S4). For the zero temperature approximation, $f(E)=1-\Theta(E-E_F)$ where $\Theta$ is a step function. Then,

\begin{align}
	\frac{I_C}{I_T}=&R\ \frac{\int_{0}^{\infty}dx \frac{c_1}{\sigma \sqrt{2\pi }}exp\left(-\frac{x^2}{2\sigma ^2}\right)\int_{0}^{R(x)}d\rho \ 2\pi\rho \int_{E_{min}(x,\rho)}^{\infty}dE_\perp D(E_\perp)\int_{0}^{E_{max}(x,\rho)}dE_\parallel [1-\Theta(E-E_F)]}{\int_{0}^{\infty}dE_\perp D(E_\perp)\int_{0}^{\infty}dE_\parallel[\Theta(E-E_F+qV_{BE})-\Theta(E-E_F)]}
\end{align}
\\[12pt]
\noindent With the J.G.Simmons' planar tunneling theory \cite{simmons1963generalized}, we can simplify the denominator in Eq. (S5).

\begin{equation}
	\frac{I_C}{I_T}=R\ \frac{\int_{0}^{\infty}dx \frac{c_1}{\sigma \sqrt{2\pi }}exp\left(-\frac{x^2}{2\sigma ^2}\right)\int_{0}^{R(x)}d\rho \ 2\pi\rho \int_{E_{min}(x,\rho)}^{\infty}dE_\perp D(E_\perp)\int_{0}^{E_{max}(x,\rho)}dE_\parallel [1-\Theta(E-E_F)]}{\frac{\hbar^2}{2m_0(\beta s)^2}\left [ \left (\varphi -\frac{qV_{BE}}{2} \right)exp\left [ -\frac{2\beta s}{\hbar}\sqrt{2m_0 \left (\varphi -\frac{qV_{BE}}{2} \right)} \right ]-\left (\varphi +\frac{qV_{BE}}{2} \right)exp\left [ -\frac{2\beta s}{\hbar}\sqrt{2m_0 \left (\varphi +\frac{qV_{BE}}{2} \right)} \right ] \right ]}
\end{equation}
\\[12pt]
\noindent where $s$ is the thickness of tunnel oxide for the direct tunneling regime, $\varphi$ is tunnel barrier height, and $\beta=1-(qV_{BE})^2/[96(\varphi+E_F-E_\perp-qV_{BE}/2)^2]\simeq 1$ \cite{simmons1963generalized}. Here, we will denote the denominator to $N$ for simplicity.

\begin{align}
	\frac{I_C}{I_T}=\frac{R}{N} \int_{0}^{\infty}dx \frac{2\pi c_1}{\sigma \sqrt{2\pi }}exp\left(-\frac{x^2}{2\sigma ^2}\right)\int_{0}^{R(x)}d\rho & \ \rho \int_{E_{min}(x,\rho)}^{\infty}dE_\perp D(E_\perp)\int_{0}^{E_{max}(x,\rho)}dE_\parallel [1-\Theta(E-E_F)]
\end{align}
\\[12pt]
\noindent The interval, where the integrand in Eq. (S7) is not zero, is $E_\perp+E_\parallel-E_F<0$. Below the threshold voltage, $E_{min}(x,\rho)+E_\parallel-E_F$ is always positive for $E_\parallel$ from 0 to $E_{max}(x,\rho)$ since $m_0>m_t$. Thus, \textit{I\textsubscript{C} / I\textsubscript{T}} is zero in the voltage range of $V_{BE}<V_{th}(x,\rho)$ and we now concentrate on the near-threshold voltage region $0<q[V_{BE}-V_{th}(x,\rho)]<<E_F$. Then,

\begin{align}
	  \frac{I_C}{I_T}=\frac{R}{N}\ K\int_{0}^{\infty}dx \ exp\left(-\frac{x^2}{2\sigma ^2}\right) & \int_{0}^{R(x)}d\rho \ \Theta \left( V_{BE}-V_{th}(x,\rho) \right) \rho  \nonumber \\[11pt] 
\times &\left[\int_{\varepsilon(x,\rho)}^{E_F}dE_\perp D(E_\perp)\left(E_F-E_\perp \right)+\int_{E_{min}(x,\rho)}^{\varepsilon(x,\rho)}dE_\perp D(E_\perp)E_{max}(x,\rho)\right]
\end{align}
\\[12pt]
\noindent where $\varepsilon(x,\rho)=E_F-\frac{m_t}{m_0}q\left( V_{BE}-V_{th}(x,\rho)\right)$ and $K=\frac{2\pi \ c_1}{\sigma \sqrt{2\pi}}$. By using the approximation $D(E_\perp)\simeq exp[ -A(E_F+\linebreak \overline{\varphi}-E_\perp)^\frac{1}{2}]$ \cite{simmons1963generalized}, where $A=(2\beta s/\hbar)(2m_0)^\frac{1}{2}$ and $\overline{\varphi}$ is the average tunnel barrier height, we can calculate the integral in square brakets for Eq. (S8).

\begin{align}
	  \frac{I_C}{I_T}=\frac{R}{N}\ K \int_{0}^{\infty}dx \ exp\left(-\frac{x^2}{2\sigma ^2}\right)& \int_{0}^{R(x)}d\rho \ \Theta \left( V_{BE}-V_{th}(x,\rho) \right) \rho \nonumber \\[11pt]
\times &\left[\int_{\varepsilon(x,\rho)}^{E_F}dE_\perp \ exp[ -A(E_F+\overline{\varphi}-E_\perp)^\frac{1}{2}] \left(E_F+\overline{\varphi}-E_\perp \right) \right. \nonumber \\[11pt]
\left.  \right. & \left. -\overline{\varphi} \int_{\varepsilon(x,\rho)}^{E_F}dE_\perp \ exp[ -A(E_F+\overline{\varphi}-E_\perp)^\frac{1}{2}] \right. \nonumber  \\[11pt] 
\left. \right. & \left. -\frac{m_t}{m_0-m_t}\int_{E_{min}(x,\rho)}^{\varepsilon(x,\rho)}dE_\perp \ exp[ -A(E_F+\overline{\varphi}-E_\perp)^\frac{1}{2}]\left(E_F+\overline{\varphi}-E_\perp \right) \right. \nonumber  \\[11pt] 
\left.  \right. & \left.+\frac{m_t \ \overline{\varphi} }{m_0-m_t}\int_{E_{min}(x,\rho)}^{\varepsilon(x,\rho)}dE_\perp \ exp[ -A(E_F+\overline{\varphi}-E_\perp)^\frac{1}{2}] \right. \nonumber  \\[11pt] 
\left. \right. & \left.+\frac{m_t \ q(V_{BE}-V_{th}(x,\rho))}{m_0-m_t}\int_{E_{min}(x,\rho)}^{\varepsilon(x,\rho)}dE_\perp \ exp[ -A(E_F+\overline{\varphi}-E_\perp)^\frac{1}{2}] \right]
\end{align}
\\[12pt]
\noindent We set $p=(E_F+\overline{\varphi}-E_\perp)^\frac{1}{2}$. Then,
\begin{align}
	  \frac{I_C}{I_T}=\frac{R}{N}\ 2K \int_{0}^{\infty}dx \ exp\left(-\frac{x^2}{2\sigma ^2}\right)& \int_{0}^{R(x)}d\rho \ \Theta \left( V_{BE}-V_{th}(x,\rho) \right) \rho \nonumber \\[11pt] \times &\left[\int_{\sqrt{\overline{\varphi}}}^{p_0}dp \ exp[ -A\ p]\ p^3 -\overline{\varphi} \int_{\sqrt{\overline{\varphi}}}^{p_0}dp \ exp[ -A\ p]\ p \right. \nonumber \\[11pt]
 \left. \right.&\left. -\frac{m_t}{m_0-m_t}\int_{p_0}^{p_1}dp \ exp[ -A\ p]\ p^3 +\frac{m_t \ \overline{\varphi} }{m_0-m_t}\int_{p_0}^{p_1}dp \ exp[ -A\ p]\ p \right. \nonumber \\[11pt]
\left. \right.&\left.+\frac{m_t \ q(V_{BE}-V_{th}(x,\rho))}{m_0-m_t}\int_{p_0}^{p_1}dp \ exp[ -A\ p]\ p \right]
\end{align}
\noindent where $p_0=(E_F+\overline{\varphi}-\varepsilon (x,\rho))^\frac{1}{2}=[\overline{\varphi}+\frac{m_t}{m_0}q(V_{BE}-V_{th}(x,\rho))]^\frac{1}{2}$ and $p_1=[\overline{\varphi}+q(V_{BE}-V_{th}(x,\rho))]^\frac{1}{2}$. We have $\int p \ e^{-Ap}dp = -e^{-Ap}\left[ \frac{p}{A}+\frac{1}{A^2} \right]+C$  and $\int p^3 e^{-Ap} dp = -e^{-Ap}\left[\frac{p^3}{A}+\frac{3p^2}{A^2}+\frac{6p}{A^3}+\frac{6}{A^4} \right]+C$. Thus,

\begin{align}
\nonumber \frac{I_C}{I_T}&=\frac{R}{N}\frac{4K\ \overline{\varphi}\ e^{-A\sqrt{\overline{\varphi}}}}{A^2} \int_{0}^{\infty}dx \ exp\left(-\frac{x^2}{2\sigma ^2}\right)\int_{0}^{R(x)}d\rho \ \Theta \left( V_{BE}-V_{th}(x,\rho) \right) \rho \\[11pt]
\nonumber &\hspace{2pc} \times \left[\Big(1+\frac{3}{A\sqrt{\overline{\varphi}}}+\frac{3}{A^2\overline{\varphi}}\Big)-\frac{m_t}{m_0-m_t}e^{-A(p_0-\sqrt{\overline{\varphi}})} \left(\frac{m_0}{m_t}+\frac{q[V_{BE}-V_{th}(x,\rho)]}{\overline{\varphi}}+\frac{3m_0}{m_t}\Big[\frac{p_0}{A\overline{\varphi}}+\frac{1}{A^2\overline{\varphi}}\Big]\right)\right. \\[11pt] \left.\right.&\hspace{7pc}\left.+\frac{m_t}{m_0-m_t}e^{-A(p_1-\sqrt{\overline{\varphi}})} \left( 1+\frac{q[V_{BE}-V_{th}(x,\rho)]}{\overline{\varphi}}+3\Big[\frac{p_1}{A\overline{\varphi}}+\frac{1}{A^2\overline{\varphi}}\Big]\right) \right]
\end{align}
\\[12pt]
\noindent In the Taylor series representation, $p_0=\sqrt{\overline{\varphi}}\bigg[1+\frac{1}{2}\frac{m_t}{m_0}\frac{q[V_{BE}-V_{th}(x,\rho)]}{\overline{\varphi}}-\frac{1}{8}\left(\frac{m_t}{m_0}\frac{q[V_{BE}-V_{th}(x,\rho)]}{\overline{\varphi}}\right)^2+\frac{1}{16}\left(\frac{m_t}{m_0}\frac{q[V_{BE}-V_{th}(x,\rho)]}{\overline{\varphi}}\right)^3+\cdots  \bigg]$, $p_1=\sqrt{\overline{\varphi}}\bigg[1+\frac{1}{2}\frac{q[V_{BE}-V_{th}(x,\rho)]}{\overline{\varphi}}-\frac{1}{8}\left(\frac{q[V_{BE}-V_{th}(x,\rho)]}{\overline{\varphi}}\right)^2 +\frac{1}{16}\left(\frac{q[V_{BE}-V_{th}(x,\rho)]}{\overline{\varphi}}\right)^3+\cdots  \bigg]$, and $e^{-At}=1-At+\frac{A^2}{2}t^2-\frac{A^3}{6}t^3+\cdots$. Then,

\begin{align}
\nonumber \frac{I_C}{I_T}&=\frac{R}{N}\frac{4K\ \overline{\varphi}\ e^{-A\sqrt{\overline{\varphi}}}}{A^2}\int_{0}^{\infty}dx \ exp\left(-\frac{x^2}{2\sigma ^2}\right) \int_{0}^{R(x)}d\rho \ \Theta \left( V-V_{th}(x,\rho) \right) \rho \Bigg[\frac{m_t}{m_0}\frac{A^2\overline{\varphi}}{8}\left(\frac{q(V-V_{th}(x,\rho))}{\overline{\varphi}}\right)^2 \\[11pt] 
&\hspace{6pc}-\frac{m_t}{m_0}\left(1+\frac{m_t}{m_0}\right)\frac{A^3\overline{\varphi}\sqrt{\overline{\varphi}}}{48}\left(\frac{q(V-V_{th}(x,\rho))}{\overline{\varphi}}\right)^3+O\bigg(\left(\frac{q(V-V_{th}(x,\rho))}{\overline{\varphi}}\right)^4\bigg) \Bigg]
\end{align}
\\[12pt]
\noindent The third order term in Eq. (S12) is only $\leq$ 8.9\% of the second order term where $q(V_{BE}-V_{th}(x,\rho)) \leq$ 100 meV, implying that the higher oreder terms are negligibly small. Moreover, for sufficiently low SB ($q(V_{BE}-V_{th}(x,\rho)) \geq$ 100 meV), they cannot significantly alter the planar BEES spectra due to the exponential decrease of Gaussian distribution. Thus, it is sufficient to consider only up to the third order. Let $K'_1=\frac{R}{N}\frac{Ke^{-A\sqrt{\overline{\varphi}}}}{2}\frac{m_t}{m_0}q^2$ and $K'_2=-\frac{R}{N}\frac{Ke^{-A\sqrt{\overline{\varphi}}}}{12}\frac{m_t}{m_0}\left(1+\frac{m_t}{m_0}\right)\frac{A}{\sqrt{\overline{\varphi}}}q^3$. Then,

\begin{align}
\nonumber  \frac{I_C}{I_T}\simeq &K'_1\int_{0}^{\infty}dx \ exp\left(-\frac{x^2}{2\sigma ^2}\right)\int_{0}^{R(x)}d\rho \ \Theta \left( V_{BE}-V_{th}(x,\rho) \right) \rho \left[V_{BE}-V_{th}(x,\rho) \right]^2 \\[11pt]
&+K'_2 \int_{0}^{\infty}dx \ exp\left(-\frac{x^2}{2\sigma ^2}\right)\int_{0}^{R(x)}d\rho \ \Theta \left( V_{BE}-V_{th}(x,\rho) \right) \rho \left[V_{BE}-V_{th}(x,\rho) \right]^3
\end{align}
\\[12pt]
\noindent By using the following relation, 

\begin{equation}
\Theta(V_{BE}-V_{th}(x,\rho))=\lim_{\zeta \to 0+}\frac{1}{2\pi i}\int_{-\infty}^{\infty}\frac{e^{i(V_{BE}-V_{th}(x,\rho))r}}{r-i\zeta }dr
\end{equation}
\\[12pt]
\noindent Equation (S13) becomes

\begin{align}
	  \nonumber \frac{I_C}{I_T} &\simeq K'_2\lim_{\zeta \to 0+}\frac{1}{2\pi i}\int_{-\infty}^{\infty}dr\frac{1}{r-i\zeta } \int_{0}^{\infty}dx \ exp\left[-\frac{x^2}{2\sigma ^2}+i\left(V_{BE}-\Phi_B^0\ +\left(\frac{V_{bb}}{\eta}\right)^{1/3}x\right)r\right] \\[11pt]
\nonumber &\hspace{2pc} \times \int_{0}^{R(x)}d\rho \  exp\left[-i\frac{9r}{x}\left(\frac{V_{bb}}{2W}\right)^{4/3}\rho^2\right] \Bigg[ \frac{K'_1}{K'_2} \left[V_{BE}-\Phi_B^0\ +\left(\frac{V_{bb}}{\eta}\right)^{1/3}x-\frac{9}{x}\left(\frac{V_{bb}}{2W}\right)^{4/3}\rho^2 \right]^2 \\[11pt] &\hspace{16pc}+\left[V_{BE}-\Phi_B^0\ +\left(\frac{V_{bb}}{\eta}\right)^{1/3}x-\frac{9}{x}\left(\frac{V_{bb}}{2W}\right)^{4/3}\rho^2 \right]^3\Bigg] \rho
\end{align}
\\[12pt]
\noindent Let $a=i\frac{9}{x}\left(\frac{V_{bb}}{2W}\right)^{4/3}$, $b=V_{BE}-\Phi_B^0\ +\left(\frac{V_{bb}}{\eta}\right)^{1/3}x$, and $c=\frac{K'_1}{K'_2}$

\begin{align}
\nonumber \frac{I_C}{I_T}&\simeq K'_2\lim_{\zeta \to 0+}\frac{1}{2\pi i}\int_{-\infty}^{\infty}dr\frac{exp\left[i\left(V_{BE}-\Phi_B^0\right)r\right]}{r-i\zeta } \int_{0}^{\infty}dx \Bigg[ \ exp\left(-\frac{x^2}{2\sigma ^2}\right) \bigg[-\frac{c}{2ar}\left(V_{BE}-\Phi_B^0\right)^2 \\[11pt] \nonumber &\hspace{10pc} +\frac{c}{iar^2}\left(V_{BE}-\Phi_B^0\right)+\frac{c}{ar^3}-\frac{1}{2ar}\left(V_{BE}-\Phi_B^0\right)^3+\frac{3}{2iar^2}\left(V_{BE}-\Phi_B^0\right)^2\\[11pt]
\nonumber &\hspace{10pc}+\frac{3}{ar^3}\left(V_{BE}-\Phi_B^0\right)-\frac{3}{iar^4} \bigg]+exp\left[-\frac{x^2}{2\sigma ^2}+ir\left(\frac{V_{bb}}{\eta}\right)^{1/3}x\right]\bigg[\frac{cb^2}{2ar}\\[11pt] &\hspace{10pc}-\frac{cb}{iar^2}-\frac{c}{ar^3}+\frac{b^3}{2ar}-\frac{3b^2}{2iar^2}-\frac{3b}{ar^3}+\frac{3}{iar^4}\bigg]\Bigg]
\end{align}
\\[12pt]
\noindent Let $\lambda=ax=9i\left(\frac{V_{bb}}{2W}\right)^{4/3}$, $\mu=V_{BE}-\Phi_B^0$, and $\nu=\left(\frac{V_{bb}}{\eta}\right)^{1/3}$

\begin{align}
\nonumber \frac{I_C}{I_T}&= K'_2\lim_{\zeta \to 0+}\frac{1}{2\pi i}\int_{-\infty}^{\infty}dr\frac{exp\left(i\mu r\right)}{r-i\zeta }\Bigg[\left(-\frac{c\mu^2\sigma^2}{2\lambda r}+\frac{c\mu\sigma^2}{i\lambda r^2}+\frac{c\sigma^2}{\lambda r^3}-\frac{\mu^3\sigma^2}{2\lambda r}+\frac{3\mu^2\sigma^2}{2i\lambda r^2}+\frac{3\mu\sigma^2}{\lambda r^3}-\frac{3\sigma^2}{i\lambda r^4} \right)\\[11pt] \nonumber
&\hspace{6pc}+exp\left(-\frac{\sigma^2\nu^2r^2}{2}\right) \int_{0}^{\infty} dx \ exp\left[-\frac{1}{2\sigma^2}\left(x-i\sigma^2\nu r\right)^2\right]\bigg[\frac{cx(\mu+\nu x)^2}{2\lambda r}-\frac{cx(\mu+\nu x)}{i\lambda r^2} \\[11pt] &\hspace{9pc}-\frac{cx}{\lambda r^3}+\frac{x(\mu+\nu x)^3}{2\lambda r}-\frac{3x(\mu+\nu x)^2}{2i\lambda r^2}-\frac{3x(\mu+\nu x)}{\lambda r^3}+\frac{3x}{i\lambda r^4} \bigg] \Bigg]
\end{align}
\\[12pt]
\noindent The functions $f_1(z)=zexp\left[-\frac{1}{2\sigma^2}(z-i\sigma^2\nu r)^2\right]=(x+iy)exp\left[-\frac{1}{2\sigma^2}(x+iy-iY)^2\right]$ where $Y$ is an arbitrary real number, $f_2(z)=(\nu z+\mu)f_1(z)$, $f_3(z)=(\nu z+\mu)^2f_1(z)$, and $f_4(z)=(\nu z+\mu)^3f_1(z)$ are analytic in entire complex plane $\mathbb{C}$ since they satisfy the Cauchy-Riemann equations and the first-order partial derivatives of their imagnary and real part with respect to $x$ and $y$ exist and continuous everywhere. Thus, in accordance with the Cauchy-Goursat theorem, the integral $\int_{C_1} (a_1f_1+a_2f_2+a_3f_3+a_4f_4)dz$ is zero for the contour $C_1$ where $a_1, a_2, a_3$ and $a_4$ are arbitrary complex numbers (see Fig. S3). On the other hand, 
\begin{figure}[t] 
\begin{center}
\includegraphics[width=0.6\linewidth]{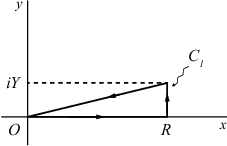}
\end{center}
\caption{Contour $C_1$}
\end{figure}
\begin{align}
 \nonumber &\int_{C_1} \left(a_1f_1(z)+a_2f_2(z)+a_3f_3(z)+a_4f_4(z)\right)dz=0 \\[11pt]\nonumber&=\int_{0}^{R} \left(a_1f_1(x)+a_2f_2(x)+a_3f_3(x)+a_4f_4(z)\right)dx+\int_{R}^{R+iY} \left(a_1f_1(z)+a_2f_2(z)+a_3f_3(z)+a_4f_4(z)\right)dz\\[11pt]
&+\int_{R+iY}^{0} \left(a_1f_1\left(x+i\frac{Y}{R}x\right)+a_2f_2\left(x+i\frac{Y}{R}x\right)+a_3f_3\left(x+i\frac{Y}{R}x\right)+a_4f_4\left(x+i\frac{Y}{R}x\right)\right)\left(1+i\frac{Y}{R}\right)dx
\end{align}
\\[12pt]
\noindent Now, we evaluate an upper bound for the integral $\int_{R}^{R+iY} \left(a_1f_1(z)+a_2f_2(z)+a_3f_3(z)+a_4f_4(z)\right)dz$. 

\begin{align*}
\nonumber &\left| \int_{R}^{R+iY} \left(a_1f_1(z)+a_2f_2(z)+a_3f_3(z)+a_4f_4(z)\right)dz \right| \\[11pt]&=\left| \int_{0}^{Y} \left(a_1f_1( R+iy)+a_2f_2( R+iy)+a_3f_3( R+iy)+a_4f_4( R+iy)\right)idy \right|\\[11pt]
\nonumber &\leq \int_{0}^{Y} \left(\left|a_1\right|+\sqrt{(\nu R+\mu)^2+\nu^2y^2}\left|a_2\right|+\left((\nu R+\mu)^2+\nu^2y^2\right)\left|a_3\right|+\left((\nu R+\mu)^2+\nu^2y^2\right)^{3/2}\left|a_4\right|\right)\\[11pt]\nonumber &\hspace{20pc}\times \sqrt{R^2+y^2}exp\left(-\frac{1}{2\sigma^2}(R^2-(y-Y)^2)\right)dy\\[11pt]
&=A'R^4exp\left(-\frac{R^2}{2\sigma^2}\right)
\end{align*}
\\[12pt]
\noindent  where $A'=\int_{0}^{Y} \big(\left|a_1\right|/R^3+\sqrt{(\nu+\mu/R)^2+\nu^2(y/R)^2}(\left|a_2\right|/R^2)+\big((\nu+\mu/R)^2+\nu^2(y/R)^2\big)(\left|a_3\right|/R)+\left((\nu+\mu/R)^2+\nu^2(y/R)^2\right)^{3/2}\left|a_4\right|\big)\sqrt{1+(y/R)^2}exp\left(\frac{1}{2\sigma^2}(y-Y)^2)\right)dy$. Consequently, Eq. (S18) becomes 
 
\begin{align}
\nonumber &\lim_{R \to \infty}\int_{0}^{R} \left(a_1f_1(x)+a_2f_2(x)+a_3f_3(x)+a_4f_4(x)\right)dx\\[22pt]\nonumber&=\lim_{R \to \infty}\int_{0}^{R+iY} \left(a_1f_1(x)+a_2f_2(x)+a_3f_3(x)+a_4f_4(x)\right)dx\\[22pt]
\nonumber&=\lim_{R \to \infty}\int_{-iY}^{R} \left(a_1f_1(x+iY)+a_2f_2(x+iY)+a_3f_3(x+iY)+a_4f_4(x+iY)\right)dx
\end{align}
\\[12pt]
\noindent As a results, Eq. (S17) becomes

\begin{align}
\nonumber \frac{I_C}{I_T}&= K'_2\lim_{\zeta \to 0+}\frac{1}{2\pi i}\int_{-\infty}^{\infty}dr\frac{exp\left(i\mu r\right)}{r-i\zeta }\Bigg[\left(-\frac{c\mu^2\sigma^2}{2\lambda r}+\frac{c\mu\sigma^2}{i\lambda r^2}+\frac{c\sigma^2}{\lambda r^3}-\frac{\mu^3\sigma^2}{2\lambda r}+\frac{3\mu^2\sigma^2}{2i\lambda r^2}+\frac{3\mu\sigma^2}{\lambda r^3}-\frac{3\sigma^2}{i\lambda r^4} \right)\\[11pt] \nonumber
&\hspace{9pc}+exp\left(-\frac{\sigma^2\nu^2r^2}{2}\right) \int_{-iY}^{\infty} ds \ exp\left(-\frac{s^2}{2\sigma^2}\right)\bigg(\frac{c\left(s+iY \right)}{2\lambda r}\left(\mu+iY\nu+\nu s\right)^2 \\[11pt] \nonumber&\hspace{9pc}-\frac{c\left(s+iY \right)}{i\lambda r^2}\left(\mu+iY\nu+\nu s \right)-\frac{c\left(s+iY \right)}{\lambda r^3}+\frac{\left(s+iY \right)}{2\lambda r}\left(\mu+iY\nu+\nu s\right)^3\\[11pt]&\hspace{7pc}-\frac{3\left(s+iY \right)}{2i\lambda r^2}\left(\mu+iY\nu+\nu s \right)^2 -\frac{3\left(s+iY \right)}{\lambda r^3}\left(\mu+iY\nu+\nu s \right)+\frac{3\left(s+iY \right)}{i\lambda r^4}\bigg) \Bigg]
\end{align}
\\[12pt]
\noindent We can use the following integral calculation results: $\int_{-iY}^{\infty} exp\left(-\frac{s^2}{2\sigma^2}\right)s^4ds=\frac{3\sqrt{2\pi}\sigma^5}{2}\Big(erf\Big(\frac{iY}{\sqrt{2}\sigma}\Big)+1\Big)-iY\sigma^4exp\Big(\frac{\sigma^2\nu^2r^2}{2}\Big)\Big(-\sigma^2\nu^2r^2+3\Big)$, $\int_{-iY}^{\infty} exp\left(-\frac{s^2}{2\sigma^2}\right)s^3ds=exp\left(\frac{\sigma^2\nu^2r^2}{2}\right)\sigma^2\left[-\sigma^4\nu^2r^2+2\sigma^2\right]$, $\int_{-iY}^{\infty} exp\Big(-\frac{s^2}{2\sigma^2}\Big)s^2ds=\frac{\sigma^2}{2}\big[\sigma\sqrt{2\pi}\Big(erf\left(\frac{iY}{\sqrt{2}\sigma}\right)+1\Big)-2iYexp\left(\frac{\sigma^2\nu^2r^2}{2}\right)\big]$, $\int exp\left(-\frac{s^2}{2\sigma^2}\right)s\ ds= \sigma^2exp\left(\frac{\sigma^2\nu^2r^2}{2}\right)$, and $\int exp\left(-\frac{s^2}{2\sigma^2}\right)ds= \frac{\sigma\sqrt{2\pi}}{2}\left[erf\left(\frac{iY}{\sqrt{2}\sigma}\right)+1\right]$. Then,

\begin{align}
\nonumber \frac{I_C}{I_T}&= \frac{K'_2}{\lambda}\lim_{\zeta \to 0+}\frac{1}{2\pi i}\int_{-\infty}^{\infty}dr\frac{exp\left(i\mu r\right)}{r-i\zeta }\Bigg[i\sigma^6\nu^4+\frac{c(2i\mu\sigma^4\nu^2-\sigma^6\nu^4r)}{2}+\frac{3i\mu^2\sigma^4\nu^2-3\mu^2\sigma^6\nu^4r-i\sigma^8\nu^6r^2}{2}\\[11pt]\nonumber&\hspace{9pc}+\frac{\sigma\sqrt{2\pi}}{2}exp\left(-\frac{\sigma^2\nu^2r^2}{2}\right)\Big(erf\Big(\frac{i\sigma\nu r}{\sqrt{2}}\Big)+1\Big)\bigg[\frac{ci\sigma^4\nu^3}{2}+\frac{3i\sigma^4\nu^3(\mu+i\sigma^2\nu^2r)}{2}\\[11pt]&\hspace{9pc}+\frac{ci\sigma^2\nu(\mu+i\sigma^2\nu^2r)^2}{2}+\frac{i\sigma^2\nu(\mu+i\sigma^2\nu^2r)^3}{2}\bigg]\Bigg]
\end{align}
\\[12pt]
\noindent The function $g(z)=exp\left(i\mu r\right)\Big[i\sigma^6\nu^4+\frac{c(2i\mu\sigma^4\nu^2-\sigma^6\nu^4r)}{2}+\frac{3i\mu^2\sigma^4\nu^2-3\mu^2\sigma^6\nu^4r-i\sigma^8\nu^6r^2}{2}+\frac{\sigma\sqrt{2\pi}}{2}exp\left(-\frac{\sigma^2\nu^2r^2}{2}\right)\\ \times \Big(erf\Big(\frac{i\sigma\nu r}{\sqrt{2}}\Big)+1\Big)\Big(\frac{ci\sigma^4\nu^3}{2}+\frac{3i\sigma^4\nu^3(\mu+i\sigma^2\nu^2r)}{2}+\frac{ci\sigma^2\nu(\mu+i\sigma^2\nu^2r)^2}{2}+\frac{i\sigma^2\nu(\mu+i\sigma^2\nu^2r)^3}{2}\Big)\Big]  $ is holomorphic on contour $C_2$ (see Fig. S4). Thus, for contour $C_2$, it satisfies the following Cauchy integral formula:

\begin{align}
\int_{C_2}\ \frac{g(z)}{z-i\zeta}\ dz=2\pi i g(i\zeta)=\int_{-R}^{R}\ \frac{g(x)}{x-i\zeta}\ dx+\int_{C_R}\ \frac{g(z)}{z-i\zeta}\ dz
\end{align}
\\[12pt]
\noindent where $C_R$ is an upper half semi-circle path. Now, we evaluate the second term in Eq. (S21). For large $R \to \infty$ and small $\zeta \to 0+$, 
\begin{figure}[t] 
\begin{center}
\includegraphics[width=0.6\linewidth]{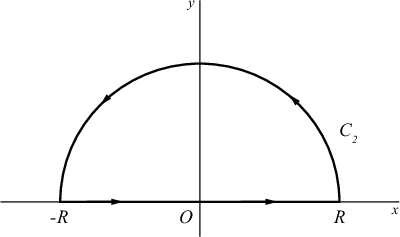}
\end{center}
\caption{Contour $C_2$}
\end{figure}

\begin{align*}
\nonumber \int_{C_R}\ \frac{g(z)}{z-i\zeta}\ dz &=i\int_{0}^{\pi}  exp\left(i\mu Re^{i\theta}\right)\Bigg[i\sigma^6\nu^4+\frac{c(2i\mu\sigma^4\nu^2-\sigma^6\nu^4Re^{i\theta})}{2}\\[11pt]&\hspace{9pc}+\frac{3i\mu^2\sigma^4\nu^2-3\mu^2\sigma^6\nu^4Re^{i\theta}-i\sigma^8\nu^6R^2e^{2i\theta}}{2}\\[11pt]\nonumber &\hspace{9pc}+\frac{\sigma\sqrt{2\pi}}{2}exp\left(-\frac{\sigma^2\nu^2R^2e^{2i\theta}}{2}\right)\Big(erf\Big(\frac{i\sigma\nu Re^{i\theta}}{\sqrt{2}}\Big)+1\Big)\\[11pt]\nonumber&\hspace{9pc}\times \bigg[\frac{ci\sigma^4\nu^3}{2}+\frac{3i\sigma^4\nu^3(\mu+i\sigma^2\nu^2Re^{i\theta})}{2}+\frac{ci\sigma^2\nu(\mu+i\sigma^2\nu^2Re^{i\theta})^2}{2}\\[11pt]&\hspace{9pc}+\frac{i\sigma^2\nu(\mu+i\sigma^2\nu^2Re^{i\theta})^3}{2}\bigg]\Bigg]\frac{Re^{i\theta}}{Re^{i\theta}-i\zeta} d\theta \\[22pt]\nonumber &\simeq i\int_{0}^{\pi}  exp\left(i\mu Re^{i\theta}\right)\Bigg[-\frac{i\sigma^8\nu^6R^2e^{2i\theta}}{2}\\[11pt]\nonumber &\hspace{9pc}+\frac{\sigma^9\nu^7\sqrt{2\pi}R^3e^{3i\theta}}{4}exp\left(-\frac{\sigma^2\nu^2R^2e^{2i\theta}}{2}\right)\Big(erf\Big(\frac{i\sigma\nu Re^{i\theta}}{\sqrt{2}}\Big)+1\Big)\Bigg] d\theta
\end{align*}
\\[12pt]
\noindent By using the asymptotic expansion for the error function $erf(z)\sim 1-\frac{e^{-z^2}}{z\sqrt{\pi}}$ for $z\gg1$,

\begin{align*}
\nonumber \int_{C_R}\ \frac{g(z)}{z-i\zeta}\ dz & \simeq i\int_{0}^{\pi}  exp\left(i\mu Re^{i\theta}\right)\Bigg[-\frac{i\sigma^8\nu^6R^2e^{2i\theta}}{2}+\frac{\sigma^9\nu^7\sqrt{2\pi}R^3e^{3i\theta}}{4}\\[11pt]\nonumber &\hspace{9pc}\times exp\left(-\frac{\sigma^2\nu^2R^2e^{2i\theta}}{2}\right)\Bigg(2-\sqrt{\frac{2}{\pi}}\frac{exp\left(\frac{\sigma^2\nu^2R^2e^{2i\theta}}{2}\right)}{i\sigma\nu Re^{i\theta}}\Bigg)\Bigg] d\theta\\[22pt]
&=i\int_{0}^{\pi}  exp\left(i\mu Re^{i\theta}\right)\Bigg[\frac{\sigma^9\nu^7\sqrt{2\pi}R^3e^{3i\theta}}{2}exp\left(-\frac{\sigma^2\nu^2R^2e^{2i\theta}}{2}\right)\Bigg] d\theta \\[22pt]
& \simeq \frac{i\sigma^9\nu^7\sqrt{2\pi}}{2} \int_{0}^{\pi} R^3e^{3i\theta}exp\left(-\frac{\sigma^2\nu^2R^2e^{2i\theta}}{2}\right) d\theta \\[22pt]
& = \frac{\sigma^9\nu^7\sqrt{2\pi}}{2} \int_{1}^{-1} R^3u^2exp\left(-\frac{\sigma^2\nu^2R^2u^2}{2}\right) du \\[22pt]
& \simeq 0
\end{align*}
\\[12pt]
\noindent Thus, 

\begin{align*}
\lim_{\zeta \to 0+}\frac{1}{2\pi i}\int_{-\infty}^{\infty}\ \frac{g(x)}{x-i\zeta}\ dx=\lim_{\zeta \to 0+} \ g(i\zeta)= \frac{\sqrt{2\pi}}{2}\frac{i\sigma^3\nu}{2}\bigg[\sqrt{\frac{2}{\pi}}2\sigma^3\nu^3+\sqrt{\frac{2}{\pi}}2c\mu\sigma\nu+\sqrt{\frac{2}{\pi}}3\mu^2\sigma\nu+c\sigma^2\nu^2\\[11pt]\hspace{11pc}+3\mu\sigma^2\nu^2+c\mu^2+\mu^3\bigg]
\end{align*}
\\[12pt]
\noindent Therefore, the planar BEES signal $I_C/I_T$ contributed by low SB patches becomes

\begin{align}
\nonumber\frac{I_C}{I_T}&\simeq \frac{K'_2}{\lambda}\lim_{\zeta \to 0+}\frac{1}{2\pi i}\int_{-\infty}^{\infty}\ \frac{g(x)}{x-i\zeta}\ dx\\[22pt]
\nonumber &=K''C_p\left[-\sqrt{\frac{2}{\pi}}2\sigma^3\nu^3\frac{1}{|c|}+\sqrt{\frac{2}{\pi}}2\mu\sigma\nu-\sqrt{\frac{2}{\pi}}3\mu^2\sigma\nu\frac{1}{|c|}+\sigma^2\nu^2-3\mu\sigma^2\nu^2\frac{1}{|c|}+\mu^2-\mu^3\frac{1}{|c|}\right]
\end{align}
\\[12pt]
\noindent where $K''=\frac{R}{N}\frac{e^{-A\sqrt{\overline{\varphi}}}}{2}\frac{m_t}{m_0}q^2$ and $|c|=6\sqrt{\overline{\varphi}}/[qA\big(1+\frac{m_t}{m_0}\big)]$. By using the mathematical nature of the threshold, $d(I_C/I_T)/dV_{BE}|_{V_{BE}=V_{th}^{eff}}=0$ and $d^2(I_C/I_T)/dV_{BE}^2|_{V_{BE}=V_{th}^{eff}}>0$, we can find $V_{th}^{eff}$ of $I_C/I_T$ for the inhomogeneous MS interface. 

\begin{align}
\nonumber \frac{d(I_C/I_T)}{dV_{BE}}\bigg|_{V_{BE}=V_{th}^{eff}}=0&=-3\mu^2-2\left(\sqrt{\frac{2}{\pi}}3\sigma\nu-|c|\right)\mu+\sqrt{\frac{2}{\pi}}2|c|\sigma\nu-3\sigma^2\nu^2\bigg|_{V_{BE}=V_{th}^{eff}}\\[22pt]
\nonumber \Rightarrow \ \mu\big|_{V_{BE}=V_{th}^{eff}}&=\frac{-\left(\sqrt{\frac{2}{\pi}}3\sigma\nu-|c|\right)-\sqrt{\left(\sqrt{\frac{2}{\pi}}3\sigma\nu-|c|\right)^2-3\left(-\sqrt{\frac{2}{\pi}}2|c|\sigma\nu+3\sigma^2\nu^2\right)}}{3}\\[22pt] \nonumber
V_{th}^{eff}&=\Phi_B^0-\sqrt{\frac{2}{\pi}}\left(\frac{V_{bb}}{\eta}\right)^{1/3}\sigma+\frac{|c|-\sqrt{|c|^2-9\Big(1-\frac{2}{\pi}\Big)\left(\frac{V_{bb}}{\eta}\right)^{2/3}\sigma^2}}{3}\\[22pt]&\simeq\Phi_B^0-\sqrt{\frac{2}{\pi}}\left(\frac{V_{bb}}{\eta}\right)^{1/3}\sigma
\end{align}
\\[12pt]
\noindent Our theory recovers to the Bell-Kaiser theory in the limit of no barrier inhomogeneity ($\sigma \to 0$). For low barrier inhomogeneity (sufficiently small $\sigma$), it suggests that the $V_{th}^{eff}$ follows a 1/3-power law dependence on $V_{CB}$, rather than the 1/4-power law predicted by the image force lowering.

\end{document}